# The role of β-titanium ligaments in the deformation of dual phase titanium alloys

Tea-Sung Jun[1,2], Xavier Maeder[3], Ayan Bhowmik[1,a], Gaylord Guillonneau[3,4], Johann Michler[3], Finn Giuliani[1], T. Ben Britton[1*]

[1]Department of Materials, Royal School of Mines, Imperial College London, London SW7 2AZ, UK
[2]Department of Mechanical Engineering, Incheon National University, Incheon 22012, Republic of Korea
[3]EMPA, Swiss Federal Laboratories for Materials Science and Technology, Laboratory for Mechanics of Materials and Nanostructures, Feuerwerkerstrasse 39, CH-3602 Thun, Switzerland
[4]Université de Lyon, Ecole Centrale de Lyon, LTDS UMR CNRS 5513, 36 Avenue Guy de Collongue, 69134 Ecully Cedex, France
[a]now at Rolls-Royce@NTU Corporate Lab, Nanyang Technological University, Singapore

[*]Corresponding author. Email address: *b.britton@imperial.ac.uk*

**Abstract**

Multiphase titanium alloys are critical materials in high value engineering components, for instance in aero engines. Microstructural complexity is exploited through interface engineering during mechanical processing to realise significant improvements in fatigue and fracture resistance and strength. In this work, we explore the role of select interfaces using in-situ micromechanical testing with concurrent observations from high angular resolution electron backscatter diffraction (HR-EBSD). Our results are supported with post mortem transmission electron microscopy (TEM). Using micro-pillar compression, we performed in-depth analysis of the role of select β-titanium (body centred cubic) ligaments which separate neighbouring α-titanium (hexagonal close packed) regions and inhibit the dislocation motion and impact strength during mechanical deformation. These results shed light on the strengthening mechanisms and those that can lead to strain localisation during fatigue and failure.

## 1. Introduction

Demands for using dual-phase titanium alloys have progressively increased in many industries, particularly for elevated temperature applications such as gas turbine and aerostructures. In these applications the microstructures of these alloys are carefully designed by means of several parameters (e.g. chemical composition, thermomechanical processing route, and ultimately microstructure) to achieving target properties. For demanding load regimes found in service it is useful to understand the role of each microstructural unit and how they interact. This is complex to interpret in a bulk polycrystal and therefore isolation of these units and testing using micro-size samples is attractive.

We have a significant interest in fatigue, and in particular the cold dwell fatigue in titanium alloys. This is a process where it is thought that microstructure heterogeneity can result in the amplification of stresses at a 'hard soft' interface for a rogue grain combination during a load hold [1-4]. The hard grain is aligned with the <c> axis of the hexagonal unit cell aligned along the axis of maximum principal stress and the soft grain has the <c> axis perpendicular to this. Through consideration of the Stroh model[5], Dunne et al.[2] suggest that under load-hold during a dwell cycle the stress within the hard grain increases as a result of dislocation based creep (even at room temperature) within the soft grain and this is supported through evidence of the quasi brittle facets formed within the hard grains as observed by and Bache et al.[6]. Combability at the interface results in an increase in stress resolved onto the basal plane of the hard grain and this can lead to facet nucleation and cracking of the microstructural unit. This is thought to lead to the cold dwell fatigue "dwell debit", where the presence of a hold at maximum load can result in a significant reduction in the number of cycles to failure in some, but interestingly not all, titanium alloys [7].

The rogue grain combination is articulated in the literature in terms of grains, but these are often considered in terms of aggregate microtextured regions, macrozones or effective structural units. From a 'bottom up' understanding, it is unclear how contiguous these 'grain regions' must be. Specifically, in the titanium system complexity of these grain regions is incurred due the presence of the Burgers Orientation Relationship (i.e. $\{0001\}_\alpha \parallel \{110\}_\beta$ and $\langle 11\bar{2}0\rangle_\alpha \parallel \langle 111\rangle_\beta$) between both phases, where there is a specific overlap of crystallographic directions and planes between the α-titanium (HCP) and β-titanium (BCC) due to the solid state-based phase transformation during thermomechanical processing. In this context, we are motivated to investigate how a single interface affects the pile-up of dislocations, the associated back stress, and the generation of geometrically necessary dislocation (GND) density associated with individual β-ligaments which can be found segmenting clusters of α-grains (of similar orientation).

We are not the first to investigate titanium alloys with micromechanical tests. Work by Gong and Wilkinson has shown that micro-cantilever testing can extract these individual slip strengths in both commercially pure alloys but also in Ti-6Al and in dual phase Ti-6Al-4V [8]. While the mechanical tests show that the presence of the dual phase microstructure in Ti-6Al-4V impacts some strength, exquisite STEM based observations [9] of the dislocations structures in these multiphase cantilevers reveal that the presence of a thin β phase does little to directly influence the kinematics of the deformation patterning within a single α colony (i.e. regions where neighbouring α and β phases adhere to the Burgers Orientation Relationship (BOR) [10, 11]). Furthermore, scaling up these micro-cantilever based tests to industrial problems has shown significant promise in both understanding the role of oxygen enriched layers in Ti-6Al-4V [12], but also in predicting the uniaxial flow behaviour of other hexagonal materials such as commercially pure zirconium [13].

This rich body of micro-cantilever based work hints that we can use micro-scaled mechanical testing to understand the behaviour of microstructural units. This has been built upon by Jun *et al*. [14], who used *in-situ* micropillar compression testing to investigate dual phase titanium micropillars where the role of individual microstructural morphologies had a marked effect on the mechanical behaviour. While striking differences in mechanical response were observed [14], there was limited analysis on the precise deformation mechanisms and the underlying microstructure control on the deformation behaviour. In essence this prior work [14] showed that the volume fraction and morphology of the tri-crystal (α-β-α) micropillars extracted from the same colony had different apparent yield and hardening response, similar observations can be found in the early work by Chan [15].

These micromechanical tests provide detailed understanding of microstructural units within macroscopic polycrystalline alloys. The deformation mechanisms of dual phase α/β Ti alloys have been studied at longer length (i.e. in polycrystalline samples) scales, and the role of individual microstructural units remains unclear. In addition to highly localised deformation and elastic/plastic anisotropy inherent to α-Ti (HCP), complexities arise due to the differences and interactions between α and β phases [15, 16]. A series of papers by the Mills group [16-19] have demonstrated the effect of deformation of Ti alloys with regards to Burgers Orientation Relationship (i.e. $\{0001\}_\alpha \parallel \{110\}_\beta$ and $\langle 11\bar{2}0\rangle_\alpha \parallel \langle 111\rangle_\beta$) between both phases. They observed a significant anisotropy in deformation behaviour which stemmed from the relative misalignment of Burgers vectors, such that $\langle a_1 \rangle$ and $\langle b_1 \rangle$ directions are closely aligned and $\langle a_2 \rangle$ and $\langle b_2 \rangle$ directions are the next most closely aligned. He *et al* [20] investigated the effect of retained β layer on slip transmission across α/β phase boundaries. They reported that the BOR has a significant influence on the slip transmission, but also argued that the orientation relationship is not necessarily required for slip transmission as the slip can still transmit through the β layer by dislocation motions moving through relatively easier slip systems (i.e. compatibility can be maintained through indirect slip transfer mechanisms).

Sandala [21] linked the likelihood of the slip transmission with β volume fraction and revealing that the alignment of slip systems as well as the amount of β phase (i.e. width of β) has a significant impact upon deformation, enabling or blocking of local shear.

In this study we investigate the effect of local phase morphology on deformation process of a dual phase Ti alloy. Target microstructural regions, with comparative α phase crystal orientations but different β morphologies, were identified using SEM and EBSD. Many micropillars were cut using FIB machining and two are shown in the present work which represent the data well for the other tests. These two pillars tested and during loading HR-EBSD maps were obtained. After the tests, post-mortem TEM imaging was used to understand the deformation modes further and support initial slip trace analysis. The focus of this work is on how β morphology can influence the slip activity, stress localisation and relaxation. The observed morphological effects are discussed in the context of deformation patterning within multiphase titanium alloys, with an ultimate goal of establishing understanding that helps us address cold dwell fatigue.

## 2. Experimental

## 2.1. Material preparation

A small block of 6 x 4 x 2 mm$^3$ was sectioned from a dual-phase Ti alloy, Ti6242 (Ti-6Al-2Sn-4Zr-2Mo in wt%), supplied by IMR (Institute of Metal research, China). A heat treatment was used to grow a large colony microstructure, holding at temperature of β transus + 50°C (i.e. 1040°C) for 8 hours and cooling down with a sufficiently slow rate of 1°C/min. This microstructure has colonies containing large α-lamella separated by thin β-ligaments (this is the same heat treatment employed in previous studies [14, 22]).

Both top and cross-section surface (see Figure 1) were mechanically polished with up to 4000 grit SiC carbide paper and finally polished with ~50 nm OP-S (Oxide Polishing Suspensions) diluted with H$_2$O by a ratio 1:5 of OP-S:H$_2$O. The top surface was etched for ~15 seconds by the Kroll's reagent (i.e. 2% HF, 10% HNO$_3$ and 88% H$_2$O) and imaged by optical microscope to identify the surface morphology.

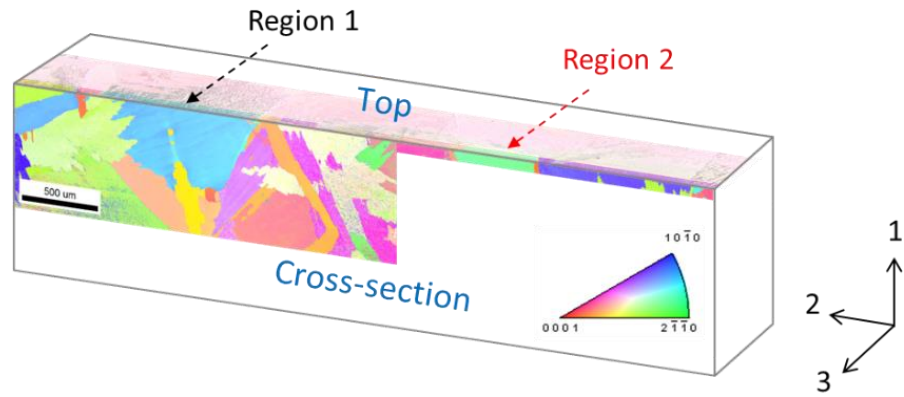

*Figure 1. Schematic diagram of a Ti6242 block: optical micrograph image with polarised light is superimposed on the top surface, and inverse pole figure (IPF) maps are also superimposed on both top and cross-section surface. These maps are used to find target grains (i.e. regions 1 and 2). Note that the IPF maps show the crystal orientations with respect to the 1-direction.*

The crystal orientations of the colonies were identified on the top and side surfaces using EBSD, in a Tescan Lyra3 FEG FIB-SEM equipped with a TSL-EDAX EBSD system. From this sample, two specific regions were selected that would trigger $\langle a \rangle$ prism slip in the α micropillars, but contain different β lamella microstructures. The pair of inverse pole figure EBSD maps are shown in Figure 1, where the grain orientations are shown with respect to the loading (i.e. 1) axis. Further higher spatial resolution EBSD maps of these two target regions were captured to reveal the β orientation.

**2.2. Micropillar fabrication**

A micropillar with a square cross-section was fabricated in each region using FIB milling. Fabrication was performed decreasing the probe current as the pillar was thinned to reduce FIB damage (i.e. surface amorphisation or Ga-contamination). Not only was this important to ensure that the mechanical response represents the underlying material, but also re-deposition and amorphisation at the free surface would reduce the quality of the surface EBSD patterns required for the *in-situ* HR-EBSD analysis. The fabricated pillars shown in Figure 2 reveal that there are several thin inclined β laths within a pillar in region 1 (i.e. β-inclined) and a thick vertical β lath in region 2 (i.e. β-vertical). The pillar dimensions were measured after each pillar was fabricated (see Table 1), and the aspect ratios are 3.2~3.5:1, which is unlikely to result buckling and result in significant strain errors [23].

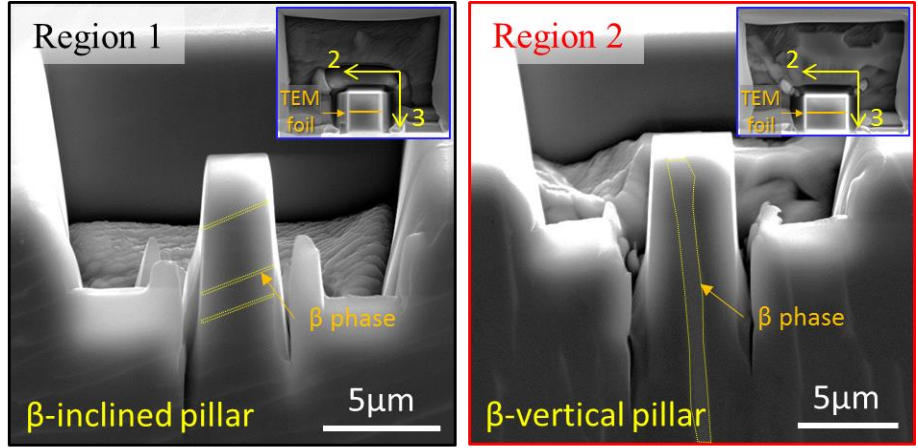

*Figure 2. SEM micrographs of square-shape micropillars fabricated in (left) region 1 and (right) region 2: The β phase has been highlighted with a yellow outline. The contrast difference clearly reveals the α (darker) and β (brighter) phases on each pillar. The β morphologies within both pillars are clearly different, such as an inclined structure in region 1 and a vertical structure in region 2. The inset shows the top view of the pillar and indicates the position that TEM foil fabricated.*

*Table 1. Dimensions of the fabricated micropillars.*

| Region | β structure | Taper angle(°) | Area@top (µm$^2$) | Height (µm) | Aspect ratio |
|---|---|---|---|---|---|
| 1 | inclined | 2.5 | 2.8 x 3.5 | 10.0 | ~3.2:1 |
| 2 | vertical | 1.7 | 3.9 x 3.1 | 12.4 | ~3.5:1 |

## 2.3. Micromechanical testing

Displacement controlled *in-situ* micropillar compression tests were performed at room temperature with an Alemnis nanoindentation platform set in a SEM [24, 25], as schematically shown in Figure 3. The platform has a pre-tilt angle of 30° with respect to the horizontal plane, and the Ti6242 block was carefully mounted so that the total tilt of the sampled plane was 70° (i.e. a tilt angle of -20° with respect to the electron beam path), which allows the EBSD measurement during the compression test. The micropillars were near uniaxially compressed by a boron-doped diamond flat punch tip with a diameter of 5µm at displacement rates of ~7 nm·s$^{-1}$ (for β-inclined pillar) and ~10 nm·s$^{-1}$ (for β-vertical pillar), to the peak tip displacement of 1.5 and 1.6µm respectively. This led to an approximate strain rate of 7~8 x 10$^{-4}$·s$^{-1}$, assuming a uniform cross section. Note that a contact misalignment angle between the top surface of a pillar and the indenter tip is likely to be less than a few degrees.

The micropillar was displaced using the flat punch, and during loading the displacement was stopped for EBSD mapping of the cross section surface (i.e. the 1-2 plane shown in Figure 1 and 3) resulting a load-hold. A total of four load-holds were performed in each pillar. EBSD measurements were performed approximately thirty seconds after the holding begun, so as to minimise the load relaxation phenomenon at early holding [24].

Raw load-displacement data are presented below (Figure 4) and later the displacement data were corrected for rig compliance (Figure 7). This correction resulted in a more reasonable measurement of the elastic modulus (E) from the unloading curve [26].

Note that the secondary electron based scanning *in-situ* video was not recorded, as trial experiments revealed a build-up in carbon contamination, which worsened the EBSD pattern quality.

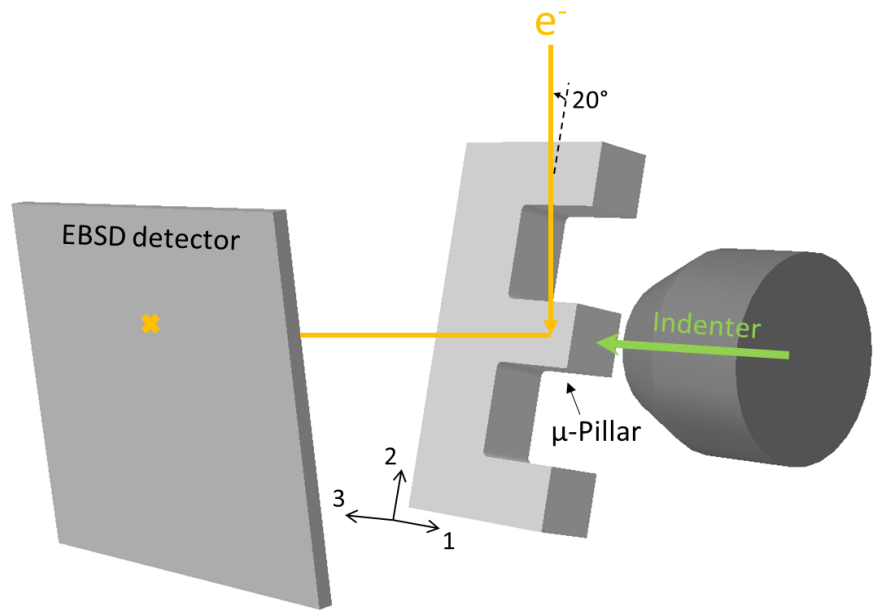

*Figure 3. Schematic of an experimental set up for the combined in-situ micropillar compression and high angular resolution electron backscatter diffraction (HR-EBSD).*

## 2.4. HR-EBSD methodology

The Kikuchi patterns were acquired to map the progress of deformation process with a step size of 200 nm and 2 x 2 binning (i.e. 512x512 pixel patterns on the Digiview camera), so as to optimise the data acquisition time and measurement sensitivity. Analysis of each map was performed offline using CrossCourt v4 from BLG productions (Bristol, UK), where threshold values above 0.04 for the mean angular error and below 0.3 for the mean cross correlation function peak height were applied to filter the data. The principle of the HR-EBSD technique is that deviatoric lattice (elastic) strains and lattice rotations result in subtle shifts of the Kikuchi patterns. These are quantitatively measured using image based cross-correlation (XCF) analysis between reference and test diffraction patterns. Sub-regions are extracted for the XCF analysis to measure the movement of zone axes across the diffraction pattern. Through tracking of four or more dispersed zone axes it is possible to extract the full (deviatoric) elastic strain tensor and lattice rotation tensor, subject to knowledge of the material elastic constants, crystal orientation (i.e. from conventional EBSD) and in assuming that the out of plane stress is equal to zero. A more detailed description can be found in the literature [27-29]. Pattern remapping was not used, as the measured lattice rotations were small.

The reference pattern is the most significant issue for determining absolute strain distributions, where in the present study it was chosen at the bottom of the pillar (i.e. substrate area) with an assumption that minimal lattice strain exists there. The same reference point was selected for each subsequent map.

GND densities were calculated from spatial derivatives of the rotation fields within CrossCourt using a L1 based line energy minimisation approach [30] including $\langle a \rangle$ basal, $\langle a \rangle$ prism, and $\langle c + a \rangle$ pyramidal dislocation types.

## 2.5. TEM investigation

Post-mortem analysis of the deformed pillars was carried out using transmission electron microscope (TEM). For this the two pillars were sectioned parallel to the loading direction with FIB-milling using Ga-ion at 30 kV ion accelerating voltage in a Helios NanoLab 600 FEG-SEM. Care was taken to contain the deformed β-lamellae from both the pillars in the sectioned foils as they were made electron transparent. The final cleaning step was done at 5 kV and an ion current of 47pA [31]. It should be noted here that fabrication of the TEM foil was rather difficult as the local residual stress induced by the mismatch of strength between the two phases led to some foil deflection. The samples were observed in a JEOL 2100F TEM using bright and dark-field techniques. In order to study the local crystallography of both the α- and β-phases, and the deformation structure, selected area diffraction patterns (SADP) were obtained under different tilt conditions.

## 3. Results

### 3.1. Load-displacement response

Figure 4 shows the significant variation in the load-displacement responses between β-inclined and β-vertical pillars. The stiffnesses ($k$) estimated from loading/unloading curves indicate that there is a micro-plasticity effect, which is likely due to a slight misalignment between the pillar and the flat punch tip, pillar geometry and local phase interactions [26]. Note here that in β-inclined pillar sudden load drops (marked by blue-dashed arrows) were caused by a combination of a relaxation during load-hold period and an external vibration from the outside of a machine.

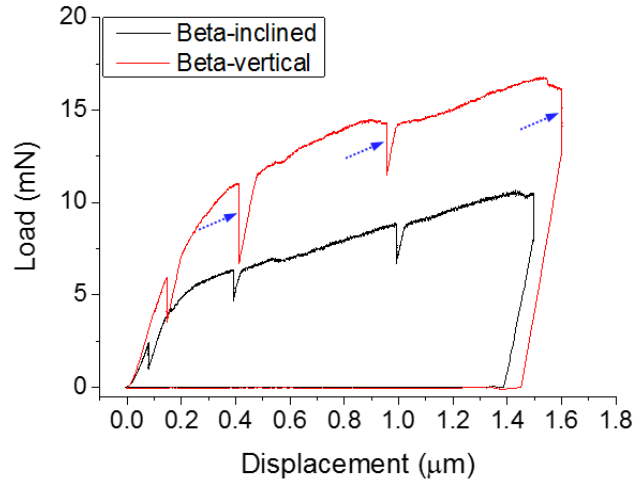

*Figure 4. Load-displacement curve observed in both β-inclined and β-vertical pillars: The blue-dashed arrows indicated sudden load drops due to an external vibration from the outside of the machine.*

### 3.2. Slip trace analysis

After the pillars were deformed, new EBSD maps and post-mortem SEM images were captured using a Carl Zeiss Auriga CrossBeam FIB-SEM with Bruker EBSD system consisting of an eFlashHR camera and Esprit v2.0 software, in order to conduct more effective slip trace analysis with pre-developed analysis methodology [14, 24]. This leads to the identification of the Euler angles, as reported using the conventions described in [32]. ($\varphi_1$, $\phi$, $\varphi_2$) such as α (212, 78, 160) and β (328, 29, 16) for β-inclined and α (221, 76, 112) and β (119, 60, 244) for β-vertical. The uncertainties associated with the Euler angle in those regions were within a range of ~2°, which is unlikely to affect identification of slip activity that gives rise to the visible slip bands.

Figure 5 displays the pole figures of α- and β-Ti observed in both regions and the corresponding unit cell structures, where the loading direction is parallel to 1-axis. The common poles overlapped in $\{0001\}_\alpha$ & $\{110\}_\beta$ and $\langle11\bar{2}0\rangle_\alpha$ & $\langle111\rangle_\beta$ indicate that the BOR is obeyed within those regions. The pole figure analysis also leads to the determination of c-, $a_1$-/$b_1$- and $a_2$-/$b_2$-directions based on the misorientations between Burgers vectors [4, 16].

The Euler angles obtained from the Bruker EBSD were then used to calculate Schmid factors (*SF*) for all 30 slip systems with a methodology described in [14]. The primary slip systems of α phase were anticipated such that $\langle a_1 \rangle$ on the prism plane (*SF* = 0.47) would be activated in the region 1 (i.e. β-inclined) and $\langle a_3 \rangle$ slip on the prism plane (*SF* = 0.46) in the region 2 (i.e. β-vertical). Schmid factors for $\langle a \rangle$-type basal and prism slips are shown in *Table 2*.

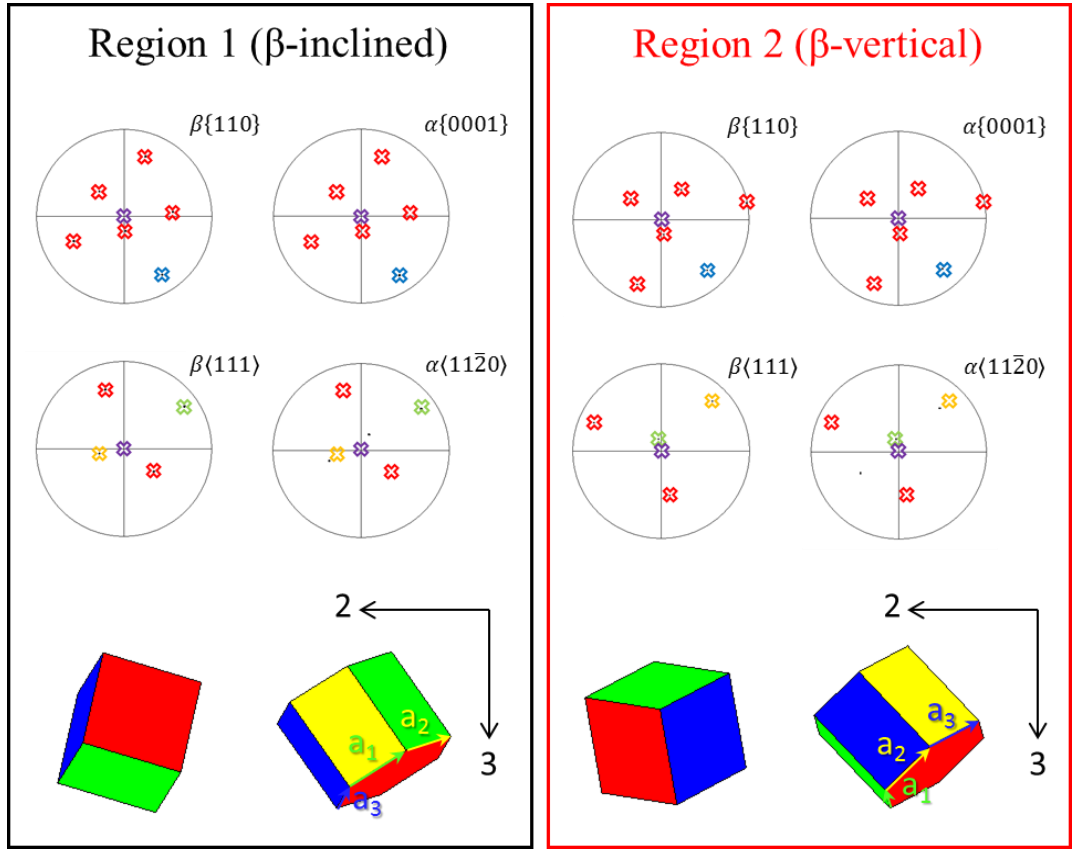

*Figure 5. Pole figure and the corresponding unit cells of α- and β-Ti in (left) region 1 and (right) region 2, indicating an obedience of Burgers orientation relationship (BOR): The purple cross represents the compression direction (i.e. along $X_1$). For the crystal plane figures, blue crosses represent the shared (0001)//(110) plane. For the crystal direction figures, the green and yellow crosses indicate a1-/b1- and a2-/b2- directions respectively. The determined ⟨a⟩ directions are superimposed in the unit cells, where $\langle a_1 \rangle$ corresponds to the well aligned Burgers vector for the BOR, $\langle a_2 \rangle$ the second most aligned and $\langle a_3 \rangle$ is the least aligned slip direction. (For interpretation of the references to colour in this figure legend, the reader is referred to the web version of this article).*

We include a full table of the Schmid factors for all the potential slip systems in the α-phase in table 2. We note that in addition to the activity of the <a> basal and <a> prism slip systems we have discussed previously, this table would indicate that alternative systems can operate. For the β-inclined pillar alternatively we could anticipate one $(0\bar{1}11)[2\bar{1}\bar{1}0]$ <a> pyramidal slip system, one $(\bar{1}101)[2\bar{1}\bar{1}3]$ <c+a> type slip system, and one <c+a> $(\bar{1}2\bar{1}2)[1\bar{2}13]$ system. The <c+a> type slip systems are unlikely to operate, due to the higher CRSS values. The <a> pyramidal slip system does not match the slip traces seen on all the faces. For the β-vertical pillar, $(1\bar{1}01)[11\bar{2}0]$ <a> pyramidal may operate and again two <c+a> type systems may operate and a similar argument can be made for the unlikely activity of these slip systems, however this does not discount screw type dislocations which could potential cross slip due to obstacles though the details of these mechanisms would be adjusted by the change in local stress tensor and not the global engineering Schmid factor type analysis presented here.

Table 2: Evaluation of the α-phase Schmid factors for the two micropillars, using the macroscopic loading state.

| | | Slip System | β-inclined | β-vertical | | Slip System | β-inclined | β-vertical |
|---|---|---|---|---|---|---|---|---|
| ⟨a⟩ Basal | ⟨a₁⟩ | (0001)[$\bar{2}$110] | 0.16 | 0.23 | | ($\bar{1}$011)[2$\bar{1}\bar{1}$3] | 0.11 | **0.48** |
| | ⟨a₂⟩ | (0001)[1$\bar{2}$10] | 0.19 | 0.09 | | (0$\bar{1}$11)[11$\bar{2}$3] | 0.06 | 0.00 |
| | ⟨a₃⟩ | (0001)[11$\bar{2}$0] | 0.04 | 0.14 | | (1$\bar{1}$01)[$\bar{1}$2$\bar{1}$3] | 0.30 | 0.00 |
| ⟨a⟩ Prism | ⟨a₁⟩ | (01$\bar{1}$0)[$\bar{2}$110] | **0.47** | 0.13 | ⟨c+a⟩ Pyram. (1ˢᵗ) | (10$\bar{1}$1)[$\bar{1}\bar{1}$23] | 0.11 | 0.11 |
| | ⟨a₂⟩ | (10$\bar{1}$0)[$\bar{1}$2$\bar{1}$0] | 0.31 | 0.33 | | (01$\bar{1}$1)[1$\bar{2}$13] | 0.30 | 0.09 |
| | ⟨a₃⟩ | ($\bar{1}$100)[$\bar{1}\bar{1}$20] | 0.17 | **0.46** | | ($\bar{1}$101)[2$\bar{1}\bar{1}$3] | **0.43** | 0.40 |
| ⟨a⟩ Pyram. | | (10$\bar{1}$1)[$\bar{1}$2$\bar{1}$0] | 0.36 | 0.33 | | ($\bar{1}$011)[11$\bar{2}$3] | 0.02 | 0.35 |
| | | (01$\bar{1}$1)[$\bar{2}$110] | 0.34 | 0.22 | | (0$\bar{1}$11)[$\bar{1}$2$\bar{1}$3] | 0.21 | 0.00 |
| | | ($\bar{1}$101)[$\bar{1}\bar{1}$20] | 0.13 | 0.34 | | (1$\bar{1}$01)[$\bar{2}$113] | 0.21 | 0.25 |
| | | ($\bar{1}$011)[1$\bar{2}$10] | 0.18 | 0.25 | | (11$\bar{2}$2)[$\bar{1}\bar{1}$23] | 0.01 | 0.08 |
| | | (0$\bar{1}$11)[2$\bar{1}\bar{1}$0] | **0.49** | 0.01 | ⟨c+a⟩ Pyram. (2ⁿᵈ) | ($\bar{1}$2$\bar{1}$2)[1$\bar{2}$13] | **0.45** | 0.07 |
| | | (1$\bar{1}$01)[11$\bar{2}$0] | 0.16 | **0.47** | | ($\bar{2}$112)[2$\bar{1}\bar{1}$3] | 0.30 | **0.49** |
| ⟨c+a⟩ Pyram. (1ˢᵗ) | | (10$\bar{1}$1)[$\bar{2}$113] | 0.09 | 0.28 | | ($\bar{1}\bar{1}$22)[11$\bar{2}$3] | 0.02 | 0.20 |
| | | (01$\bar{1}$1)[$\bar{1}\bar{1}$23] | 0.12 | 0.03 | | (1$\bar{2}$12)[$\bar{1}$2$\bar{1}$3] | 0.28 | 0.00 |
| | | ($\bar{1}$101)[1$\bar{2}$13] | 0.50 | 0.22 | | (2$\bar{1}\bar{1}$2)[$\bar{2}$113] | 0.17 | 0.30 |

Figure 6 shows the local slip activities traced on the secondary electron (SE) micrographs of deformed pillars, which were taken at the tilting angle of 70°. The unit cells superimposed are consistent with the pillar micrographs. Both pillars show planar slip behaviour which was consistently found for prism slip activity, similar to previous observations reported in [24]. In the β-inclined pillar, ⟨a₂⟩ prism slip ($SF = 0.33$, second highest among ⟨a⟩-type basal and prism slips) was primarily activated, even though ⟨a₁⟩ prism slip was anticipated as it had a higher global Schmid factor. This may be an interesting observation as the slip plane for ⟨a₁⟩ prism slip is very closely aligned with the long axis of the β lamella, and the interaction between both phases is likely to require higher energy to activate the ⟨a₁⟩ slip and consequently impede significant activity of ⟨a₁⟩ prism slip. In the β-vertical pillar, ⟨a₃⟩ prism slip was primarily activated as anticipated and further ⟨a₂⟩ prism slip ($SF = 0.31$, second highest among ⟨a⟩-type basal and prism slips) was also observed.

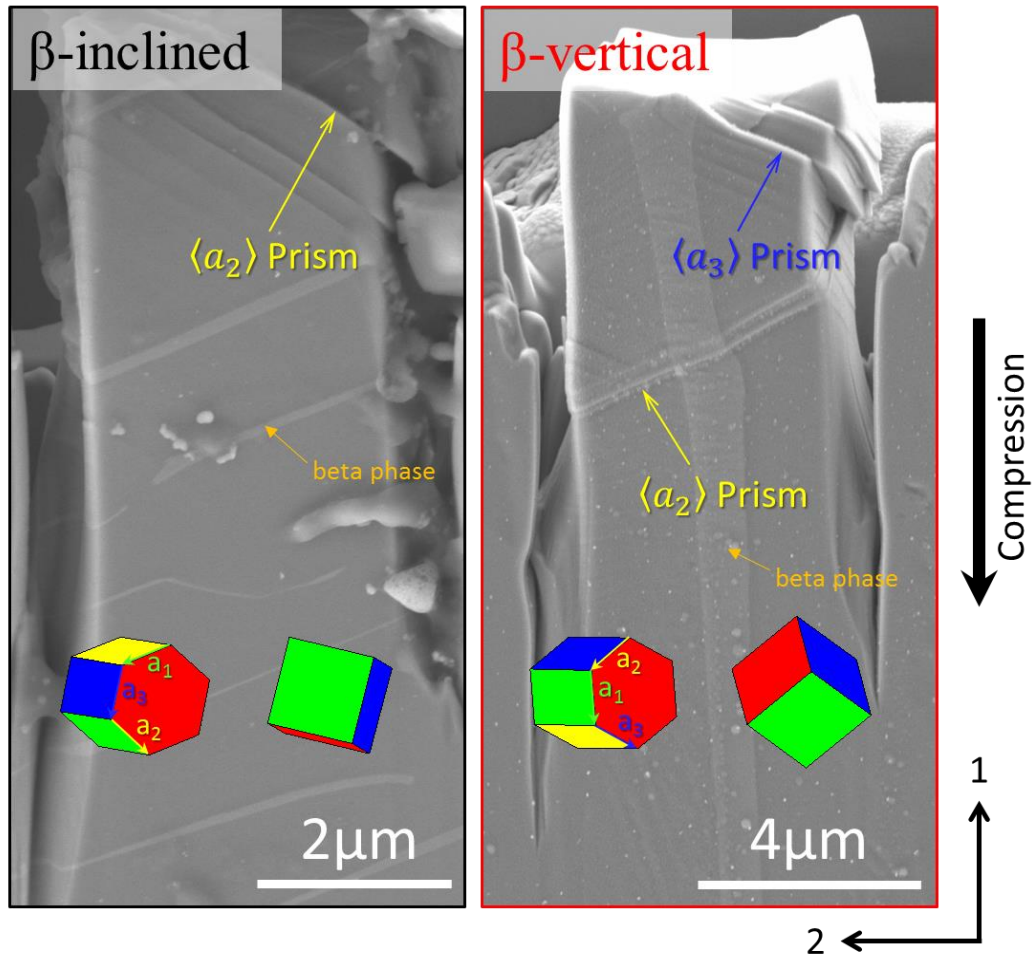

*Figure 6. Slip trace analysis of the deformed micropillars with (left) β-inclined and (right) β-vertical structure, showing the activated slip system of ⟨$a_2$⟩ prism in β-inclined and ⟨$a_2$⟩ and ⟨$a_3$⟩ prism in β-vertical.*

### 3.3. HR-EBSD analysis: stresses and GND density

Engineering stresses and strains were calculated by dividing the applied load by the cross-section area at the top of each pillar and the displacement by the height, respectively. Figure 7(A) reveals the resultant stress-strain curves, where the strain was corrected with a rig compliance correction, assuming that $E_{unloading}$ for β-inclined pillar, where the elastic modulus of α-Ti at the declination angle of ~77° is expected to be 105 GPa [33]. β-vertical pillar shows that much higher stresses are required to deform this pillar. This is likely to infer that the β phase is at least stronger than the soft orientation of α phase, although direct experimental measurements of the β properties have not been performed yet. But it is difficult to predicate that this stress difference is solely caused by the β morphology as the β volume fraction and crystal orientation also have significant effect on deformation. Further SE micrographs rotated from Figure 6 are also placed to link with the following HR-EBSD results.

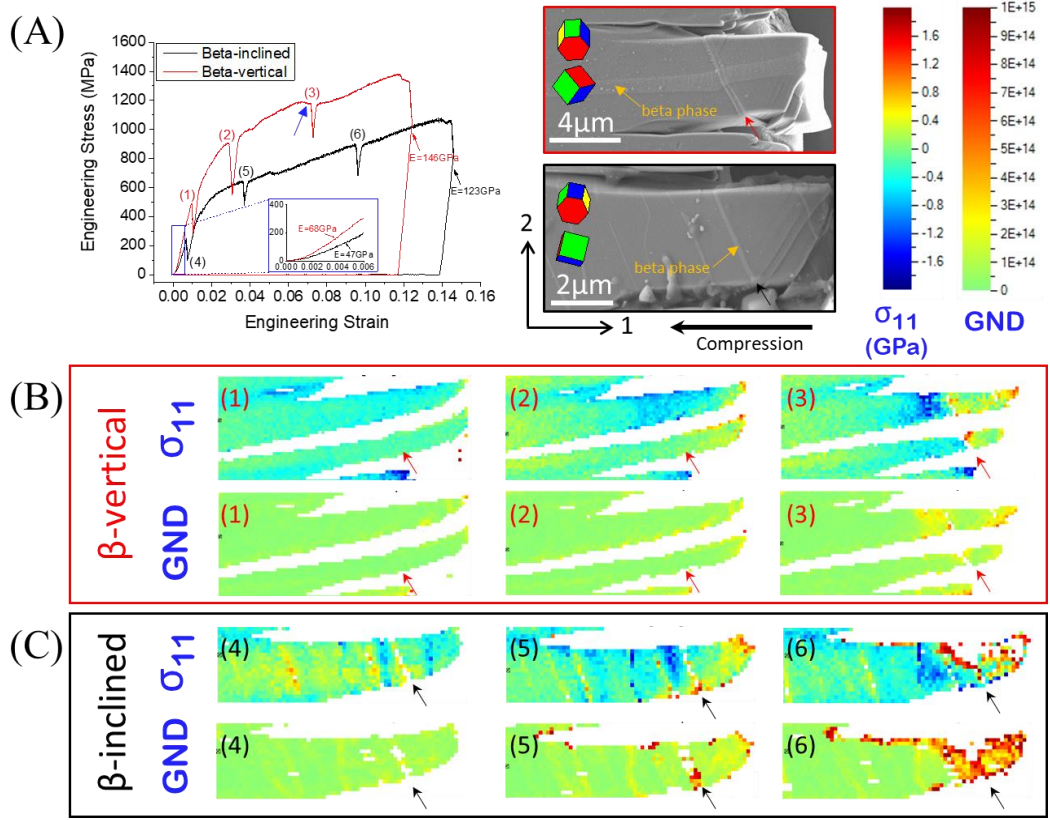

*Figure 7. Correlation between micropillar compression and HR-EBSD results: (A) Engineering stress-strain curves observed in pillars with β-inclined and β-vertical structures (shown in the right side). EBSD patterns were collected during holding periods at positions of (1) ~ (6); EBSD cross-correlation stresses calculated along the 1-axis of pillar and GND density distributions are presented in (B) for β-inclined and (C) for β-vertical. Corresponding scale bars are shown in (A). Properties from β phase were not included (i.e. remained as blank) and regions where there was no EBSD pattern have been removed through quality (peak height and mean angular error) thresholding. Positions marked on SE images by red and black arrows are consistent with those on the field plots (For interpretation of the references to colour in this figure legend, the reader is referred to the web version of this article).*

Figure 7(B) and (C) show the field plots of stress variations along the loading axis ($\sigma_{11}$) and the corresponding GND densities in β-vertical and β-inclined pillars, respectively. Note here that the insert label on each plot matches with that in the stress-strain curves. Both β structures result in quite different evolution of the stress and GND density fields.

In β-vertical pillar the $\sigma_{11}$ distributions were mostly concentrated in the mid-top area in early plastic regime (see (2), Figure 7(B)), and with further deformation split into tension and compression at near the boundary of the slip bands (see (3)). This inhomogeneity of deformation occurred as a consequence of GND evolution, where the corresponding density distribution was consistent with the evolution $\sigma_{11}$ (i.e. there is more stress supported in the hardened region). The slip band marked by red arrow shown in the micropillar image of Figure 7(B) is associated with a little stress drop in micropillar image (3), marked by blue arrow in Figure 7(A).

In β-inclined pillar, on the other hand, spatial variations in $\sigma_{11}$ develop very early (see (4) in Figure 7(C)) in the deformation series and simply strengthen in magnitude with further loading (see (5) and (6)). This is most likely as the inclined β phase results in more constrained deformation, resulting in compatibility and equilibrium constraint (even in these micropillars). This is further evidenced with the evolution of higher GND densities towards the α-β evolved at the phase boundary where $\langle a_1 \rangle$ prism slip was expected. Compressive $\sigma_{11}$ was typically found in α phase, further inferring the higher strength of β phase than that of the soft orientation of α phase.

### 3.4. TEM observation

Further investigation of the deformation structure in the micropillars with two different β orientation was carried out with respect to the α. Figure 8(A) shows the bright-field cross-sectional image (1-2 plane) of the deformed β-inclined pillar. This pillar was observed to possess three β-phase lamellae within the α colony. Note that owing to excessive thinning, the TEM foil was bent slightly. However, this did not interfere the local diffraction patterns, three of which obtained from the α-phase zone axes of $[11\bar{2}3]$, $[1\bar{2}13]$ and $[10\bar{1}2]$ are shown in Figure 8(B-D). The very distinct diffraction patterns obtained from SAD of the three sections indicates varying orientation within the foil, which has been caused by twisting of the foil during thinning. The SADP in Figure 8(C) was obtained from the α/β interface and thus has two diffraction patterns – a superposition of α and β. While the α-pattern is from $[1\bar{2}13]$ zone axis, the corresponding β pattern was from $[13\bar{1}]$ axis. The parallelism of $[1\bar{2}13]_\alpha \parallel [13\bar{1}]_\beta$ was found to satisfy the Burgers orientation relationship (this is consistent to the EBSD analysis presented in Figure 5). The majority of this α/β phase boundary looks coherent, but there was also evidence of some dislocation mediated shearing of the β lath was observed in the foil, as shown in the dotted yellow box. Here, the β lath sheared about 260 nm, which required about 900 dislocations, where the Burgers vector of $\langle a \rangle$-type slip is 0.295 nm. While this shearing directly indicates that there has been plastic slip in this region, the presence of local contrast within the foil in the α phase near the α/β interface, associated with this shearing, also supports the HR-EBSD observation of a GND field modulated by the α/β interface. However, this TEM observation is based upon an unloaded foil with significant relaxation during thinning and therefore it does not necessarily support the high concentration of stress observed at the interface resulted in slip transfer across the α/β boundary. Due to the fine structure of the β-lamellae, the defect structure within the phase could not be resolved.

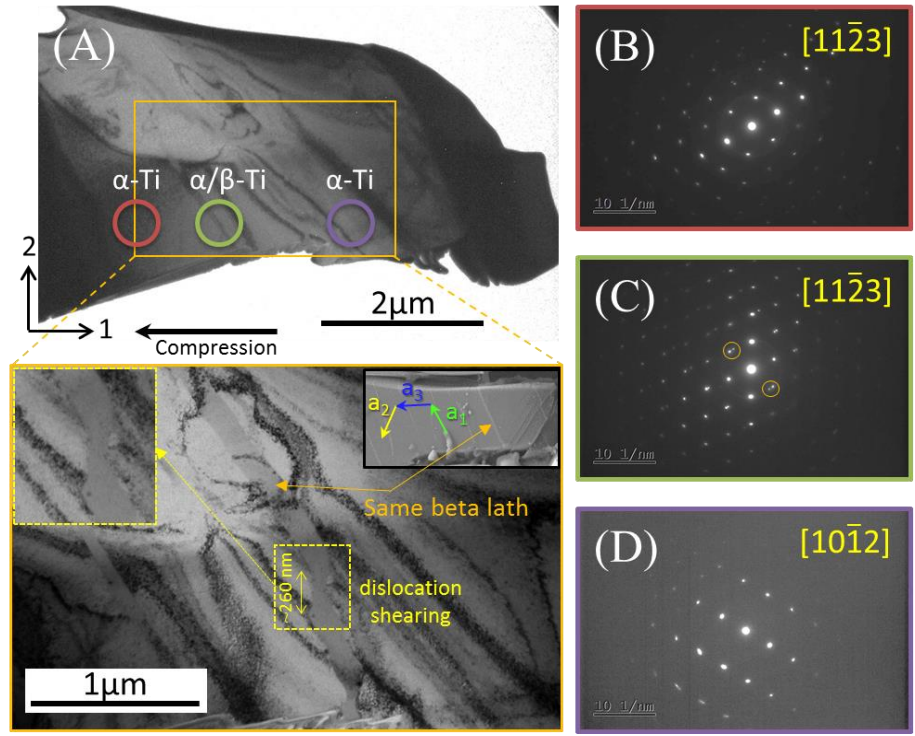

*Figure 8. (A) Cross-section TEM image of the deformed β-inclined micropillar with magnified bright field image (orange box). The α-β interface of the β-lath is sheared and this is probably due to slip activity shown in the inserted SEM image (black box); selected area diffraction patterns (SADP) obtained from different regions of the pillar marked by (B) light brown, (C) light green and (D) light purple circles shown in (A). The SADP in (C) is obtained from the α/β interface.*

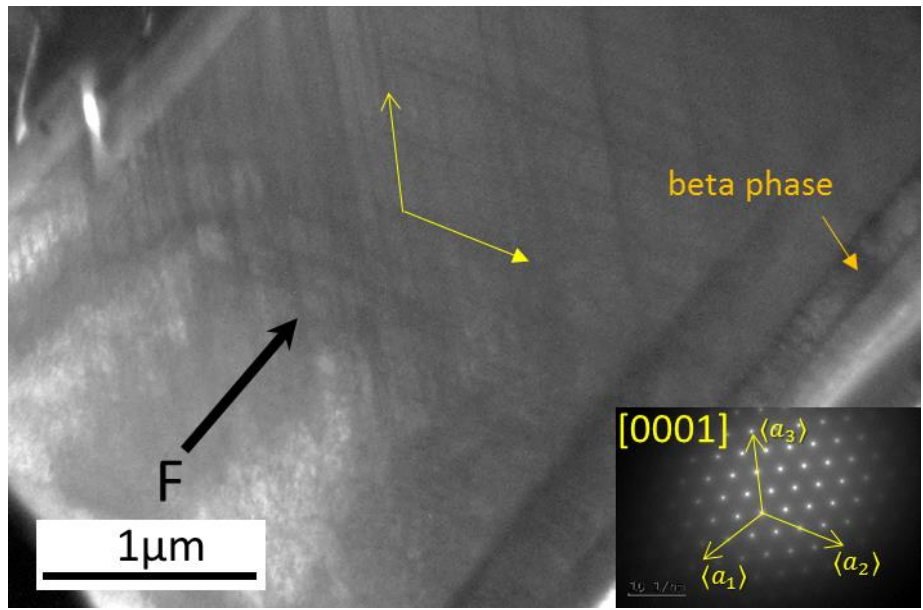

*Figure 9. Cross-section TEM image of the deformed β-vertical micropillar viewed down the ⟨c⟩-axis of the α – the three ⟨a⟩-axes on the SADP and the loading axis with respect to the*

*micropillar are marked in the respective images. A criss-cross slip pattern is activated along two prism planes, as seen on the foil surface.*

Figure 9 shows the cross-section through the β-vertical pillar. Unlike the β-inclined pillar, there was a clear formation of a slip step. Viewing down the [0001] direction of the α phase, there are high density of lines that are parallel with the $\langle a_2 \rangle$ and $\langle a_3 \rangle$ directions. Assuming the β-lamellae in the pillar lying almost parallel to the pillar axis, or the $\langle a_1 \rangle$ direction, the slip lines in the α-phase are parallel to the $\langle a_2 \rangle$ and $\langle a_3 \rangle$ directions giving rise to 120° criss-cross slip pattern over the foil surface. It can be observed that the glide step is formed parallel to the $\langle a_2 \rangle$-family of slip lines and hence clearly demonstrating slip on the prismatic plane of α. This indicates that in this orientation, the slip bands form regions of high stress concentrations, accommodating most of the applied strain until (plastic) failure of the pillar occurs by shearing along the band. The β-phase of ~0.2 µm thickness within the TEM foil section is shown in Figure 9, which is found to lie close to the edge of the pillar. No glide step was observed along that side of the pillar. In the β phase the slip lines would have the $[1\bar{1}1]$ direction with $(011)$ glide plane, but due to the small size of the β phase no internal defect structure could be resolved satisfactorily. However, the fact that no macroscopic slip occurred on the side with β-lamellae in the pillar shows that the coherent interface acts as a hindrance to the $\langle a_2 \rangle$ slip in the neighbouring α by resisting slip transfer. This is also in excellent agreement with the stress and GND maps obtained from HR-EBSD, shown in Figure 7(B), where the hotspots regions are not located at the α/β interface but rather appears as an angular band in the α-phase, which is believed to be along the $\langle a \rangle$-slip direction.

### 3.5. Strain rate sensitivity

Quantitative analysis of strain rate sensitivity was not performed here as the load-time record is likely to be influenced by both the plastic properties of a pillar and the elastic properties of testing machine, specimen and specimen mountings (e.g. glue) [34]. Instead, morphological effects on strain rate sensitivity were examined qualitatively by comparing values of $\Delta\sigma/\sigma_{t=0}$, where $\Delta\sigma$ is a stress variation between $\sigma_{t=0}$ and $\sigma_{t=300}$. Figure 10 shows the relaxation stresses of β-inclined and β-vertical pillar during the holding period, where the numbers (1)-(6) indicate the specific strains at which relaxation stresses measured (i.e. consistent with the numbers shown in Figure 7(A)). In β-vertical, values of $\Delta\sigma/\sigma_{t=0}$ were 10~11%, quite consistent through all deformation process. On the other hand, in β-inclined the value was 23% in early elastic regime, became 16% after yielding and consistent through the work hardening stage.

The variation in relaxation in these two multiphase pillars is further evidence that the morphology of the β lamella within the dual phase microstructures is of importance. In particular, load relaxation in these micropillars is likely due to the progression of mobile dislocations through a sessile dislocation structure, which evolves during the work hardening phase.

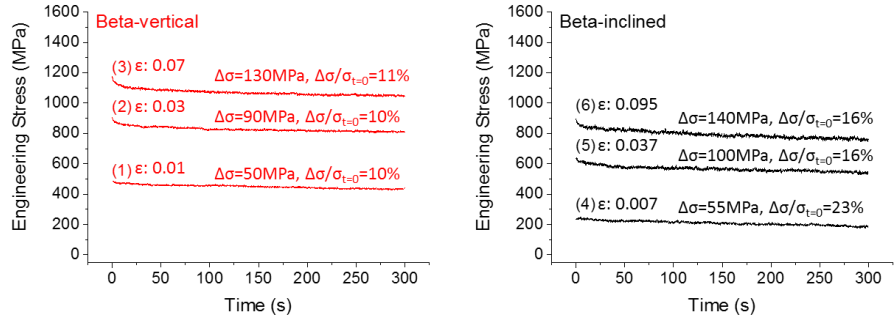

*Figure 10. Engineering stress-time curves observed in (left) β-vertical and (right) β-inclined micropillar, during the holding period (first 5 minutes).*

## 4. Discussion

In the present study we have exploited a combined *in-situ* micropillar compression and high angular resolution EBSD technique to examine the effect of β morphology in the deformation of dual phase Ti alloys, with a view to understanding individual microstructural units. Prior work has illustrated the importance of Mo in the Ti-624X alloys and this has stimulated much study on these alloys systems. Indeed, a recent simulation study has highlighted that the Mo in the α-phase is unlikely to explain Dwell susceptibility [35] and this aligns with our assumption that the subdivision of 'grains' in complex dual phase microstructures, which include the Burgers orientation relationship is critical for this phenomena. Our observations, using direct testing of individual microstructural units, reveal that deformation is patterned differently between two grains of essentially the same α-phase orientation, and same chemistry, are critical in assessing and answering: what is the "effective structural unit" [36] for these alloys systems?

We have constructed two pillars with the same nominal α-phase orientation with respect to the loading geometry. Our *in-situ* HR-EBSD observations span a range of strain states from ~1% to 10% uniaxial strain. These *in-situ* maps were taken with a 200nm resolution, which was chosen to maximise the area covered within a finite time (as each pillar compression required 5 maps to be captured) and the spatial resolution was selected to optimise recovery of the GND density, as a higher spatial resolution results in significant increase in measurement noise [28]. However, note that further quantitative interrogation of these fields is included in Figure 11. The HR-EBSD measurements provide quantitative assessment of the hardening in our pillars, rather than the slip, as there are cases where slip does not give rise to GND density [37] (nor dislocations left in our pillar for TEM analysis) but this would be evident as slip traces on the surface or significant shearing of our structure beyond what we comment on here.

To supplement our *in-situ* observations, we have performed *post mortem* TEM observations, performed at the end of our tests. These provide increased spatial resolution of the resultant dislocation structure and evidence of shear of the α/β interface and β phase, but do not have the *in-situ* quality nor quantitative lattice strain based stress measurements of the HR-EBSD data. We have used these to assist in understanding the nature of the dislocation structures contained within the deformed pillars.

Combined these two approaches provide insight into the role of interfaces in dual phase titanium alloys. Specifically that these two different interface structures during deformation perform different functions in terms of strain patterning, stress build up, and GND storage. This patterning is related to the morphology and relative crystal orientation of the two phases (note misorientation and morphology are not independent due to the precipitation and growth of α from the β during the solid state transformation [38]).

The impact of this on dwell fatigue relates to the understanding of how the stress at critical interfaces develops in the time dependant load hold where stress can increase at the interface and lead to facet formation. Therefore, our findings which include time dependant plasticity, local hardening and dislocation structure formation, provide insight into evaluation of the role of sub-structure [7].

The most dramatic observation is that macroscopic stress-strain response of the pillars is significantly different when the inclination of the β-lamella is changed. The mechanisms of this change in response has been explored using full field measurements of stress and GND content, where the β-lamella modulate the local deformation fields significantly and change the (direct or indirect) slip transfer mechanisms.

The inclination of the β lamella changes how slip occurs within the pillars and their resultant mechanical strength. In particular, it is worth noting that due to the nature of the BOR, the interface may be the $\langle a_1 \rangle$ Burgers vector. There could be an influence of the number of interfaces within the inclined β pillar, however we suspect that this is small as the back stress associated with these interfaces is smaller towards the bottom of the pillar (see Figure 7).

For the β-vertical case, this means $\langle a_1 \rangle$ slip system (which translates directly onto a slip system within the β-phase) does not have a high resolved shear stress and there is unlikely to be active. Therefore the $\langle a_2 \rangle$ and $\langle a_3 \rangle$ slip systems carry slip. The $\langle a_2 \rangle$ slip system generated the largest slip steps on the side of the pillar (the $\langle a_3 \rangle$ slip system deformed the top corner of the pillar and this slip did not penetrate the β lamella). If $\langle a_2 \rangle$ slip results in a direct slip transmission across the α/β interface, then a residual dislocation must be generated to accommodate the misalignment in Burgers vector between the two phases. Once the active slip penetrates the interface, dislocations must progress through the β phase, and again exit with local accommodation at the next α/β interface. It is likely that the progression of dislocations through the β phase and/or 'queueing' of dislocations at the interface results in the ability to withstand higher stresses. This is enhanced by the activation of $\langle a_3 \rangle$ slip and its interactions with $\langle a_2 \rangle$ mobile dislocations. The effect of this β-phase slip accommodation is borne out with increases in the local GND content and resolved $\sigma_{11}$ fields near the slip trace in the full field HR-EBSD plots (Figure 7). The second factor to consider is that this pillar consists of a near vertical sandwich of α-β-α crystals and that load shedding means that the stress could be higher due to the differences in stiffness of the α and β phases.

For the β-inclined case, the $\langle a_1 \rangle$ to be active towards the top of the pillar (Figure 6(A)). Observation of $\langle a_1 \rangle$ could be masked by slip along or near to the α/β boundaries, but the lack of changes in the surface morphology (evidenced as high contrast in the SEM) indicates that this is unlikely. Furthermore, the α/β interface has been shown to have extensive shearing by $\langle a_2 \rangle$ slip in the TEM (Figure 8). The full field stress and GND maps obtained from HR-EBSD indicate that hardening in this case is much more homogeneous and almost absent, which is supported by the lack of hardening in the macroscopic stress-strain response (Figure 7(A)).

While the inherent strength of the β phase may play a role in controlling the deformation, given that slip is seen to extend across from α to β in both cases here we have focussed on compatibility. A related study by Ashton et al. [39] has indirectly measured the importance of the β phase for vertical β ligaments sandwiched between two α laths (i.e. very similar to our vertical example here) and shown that the compatibility and crystal plasticity of the β is critical in controlling the strain rate sensitivity of a composite micropillar (similar to our observations here). While the inherent strength of the β phase may vary slightly for these two pillars, they are from the same sample and our work on these alloys shows that for this heat treatment all the β ligaments have similar chemistries. It is expected that the β ligaments here will broadly have similar mechanical responses but we cannot determine this with great precision in the absence of a large volume of β with this chemistry to create partner β pillars (we tried and failed to create homogeneous β ingots using arc melting).

In order to compare more quantitatively the stress and GND evolution of the β-vertical and β-inclined case, line scan data collected during holding periods at positions of (2-3) and (5-6) (see Figure 7) are extracted from HR-EBSD results. Figure 11 shows the $\sigma_{11}$ and GND density distributions plotted along the single line (marked by X-X' in the inserted micrograps) of each pillar. Near the yield point, stresses are rather constant in the β-vertical (2), while stress variations can be found in the β-inclined (5) particularly at the phase boundaries. The associated GND densities are such that the β-inclined has somewhat higher GND density (~9.2 × $10^{13}$ $m^{-2}$ on average) than that in the β-vertical (~4.2 × $10^{13}$ $m^{-2}$ on average) and more localised at near the phase boundaries. With further deformation, significantly higher stresses and GND densities were evolved at near the phase boundary of the β-inclined (6), where the average GND densities measured in both pillar are ~2.1 × $10^{14}$ $m^{-2}$ for the β-inclined and ~7.7 × $10^{13}$ $m^{-2}$. Note here that these results cannot be directly compared due to the difference of line scan position, phase morphology and strain levels. However, the results are clearly as a consequence of the presence and morphology of β phase, leading to the fact that the β morphological effect is likely to play a significant role in understanding the local deformation mechanism of multiphase titanium alloy.

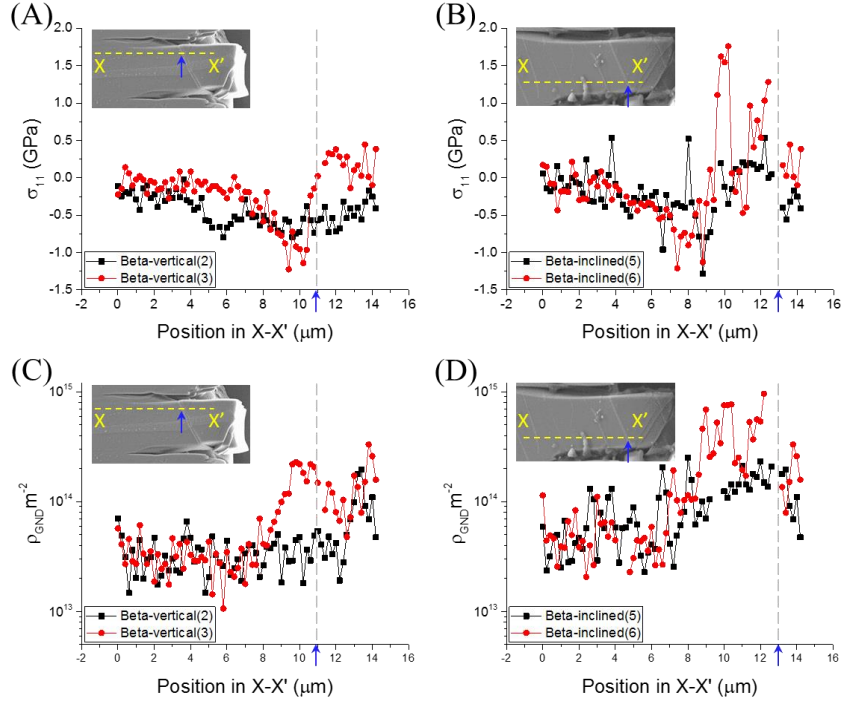

*Figure 11. Comparison of stress and GND density between the β-vertical (A and C) and the β-inclined (B and D): Line scan data are extracted from HR-EBSD results as shown in Figure 7 for maps near the initial plastic yield (2 and 5) and later in the load-displacement curve (3 and 6). Blue arrow indicates the position of slip band (for the β-vertical) and phase interface (for the β-inclined).*

The role of the slip through the β-phase is interesting. Commonly BCC metals show interesting effects associated with the progress of screw dislocations through the lattice (in extreme cases the screw dislocations are so slow that this gives rise to the ductile to brittle transition temperature). The progress of dislocations through this β-phase chemistry is unclear, and in this study the phase narrow so that it makes more detailed TEM observations challenging, especially as the role of the stress state due to the α/β interface and local dislocation content is likely to occlude definitive observations (e.g. due to bend contours). It is possible that in the vertical case, the progress of slip is severely hampered by the strength and strain rate sensitivity of the underlying β phase, almost independently of the role of the α/β interface.

As there is direct shearing of the β phase, it seems that slip within the β-phase or near the α-β interface is likely important for both cases. The effective slip length in the β-vertical case is much longer than the β-inclined case. This could be important in how the two accommodating parts of the Burgers vector are accommodated across the structure (as the $\langle a_2 \rangle$ dislocations enter and leave the lamella region there is a geometric requirement to accommodate the mismatch between the HCP Burgers vector and the BCC Burgers vectors). Controlling the microstructure experimentally thus far has proven difficult, it is likely that a study to focus on the local stress variations associated with the phase boundary accommodation of this extra Burgers vector will likely require simulation (such as discreet dislocation dynamics). Alternatively, if the stress is large enough, there may be an indirect transmission mechanisms where by dislocations are nucleated on the far side of the β-phase lamella and the exact rate sensitivity of the β-phase is less important.

The difference between the two morphologies of β lamella with a similar orientation of α colony is very interesting. In many studies, the role of the β substructure is largely ignored and the presence of an "effective structural unit" (ESU), that is a critical unit size for deformation and failure of a Ti polycrystalline aggregate [36], is thought to be tightly correlated to the prior-β grain structure / macrozone / micro-textured region size. If, as this study suggests, that the morphology of the β lamella plays a significant role in changing the slip system strength and local strain rate sensitivity, then care must be taken in extracting out intrinsic materials properties from macroscopic polycrystalline responses, as for example in this study the apparent critical resolved shear stress for each pillar (if they were considered the same pseudo-phase) would be different by a factor of 1.5 (400 vs 600 MPa). Some of this variation could be imparted due to the change in constraint associated with micropillar-testing, but the *in-situ* HR-EBSD and post-mortem TEM indicate that the mechanisms and nature of slip and hardening is also significantly changed at a local scale, and the detail of this may affect homogenisation strategies in polycrystal studies.

We urge readers to be careful when using these properties directly as there are limited experiments of the raw mechanical data presented (more substantive work can be found elsewhere [39-43]) and this work is aimed towards highlighting mechanism variation between the two morphologies. It is known that the volume fraction of the β-phase [39] and the precise structure of the β-phase within a micropillar can significantly alter the apparent strength [40]. Unfortunately nature is not kind enough to enable us to extract exact sample dimensions of our choosing with the same volume fraction for these two morphologies. In spite of these challenges, to our knowledge the present work is the first to provide substantive experimental evidence that can be used to inform higher order models that capture the role of the interfaces on the dislocation motion and relative generation of stored dislocation content and associated stress fields. Furthermore, there is a slight taper to our pillars. Such taper is intentional as it creates a minor stress gradient in the pillar, which helps stabilizing the structure during deformation (the slight lower stress at the bottom of the pillar helps stopping the defects so that they can be observed, rather than having a total breakdown of the pillar due to possible sudden dislocation burst). But this makes extraction of engineering stress-strain data difficult, and we hope the present work motivates follow-up modelling (akin to Zhang et al. [42]).

The slip trace analysis we have performed has been conducted with knowledge of the crystal orientation, loading state and BOR. It is unlikely that there are significant slip bands that are missed as this would require a slip system that has a shear direction which is entirely in plane, and alternative strain mapping techniques (e.g. total strain with DIC) are even more taxing at the lengthscale required due to tailoring of the ROI size and the fine microstructure within these pillars.

Further understanding of dwell fatigue and facet formation requires understanding of load shedding. This load shedding occurs from soft grain to neighbour hard grain and results in increased stress across grain interface (likely promoting quasi-brittle basal cleavage) and our results indicate that further stress variations are likely to be controlled by the β morphology. We note that our pillars do not expressly contain a dwell cycle, but we can explore the time constants for dwell fatigue from these types of tests. In complementary work to the present study, Zhang *et al.* [43] recently investigated the slip properties of α and β phases in Ti6242 using combined micropillar compression and computational crystal plasticity, and found that the β phase has stronger rate sensitivity than the α. Our findings, in Figure 10, hint that the presence and morphology of β phase can alter the local strain rate sensitivity, even at low strains. Our results enable us to hypothesize that the inclined β laths within a soft grain adjacent to a hard grain may generate worse scenario model associated with load shedding phenomenon, which is considered as a potential factor leading to cold dwell fatigue and ultimately the component failure. Further investigation using appropriate modelling techniques (including two phase discrete dislocation dynamics) and further experiment is likely required to confirm this hypothesis.

The volume fraction, thickness and structure of the β lamellae change with chemistry, and in particular within the Ti-624x alloys as studied extensively in the literature [22, 44-46], where x corresponds to varying Mo content. Mo is a β-stabiliser. Qiu *et al.* [46] show that changes in Mo content markedly affect the dwell sensitivity of these alloys (where Ti-6242 has been shown to be rate sensitivity and Ti-6246 is not [22]). Examining the microstructures of these alloys (Figure 3 in [46]), the primary role of changing Mo content is to change the microstructure of the alloy. This indicates that the role of the β morphology could be significant in changing dwell sensitivity, through a change in the slip behaviour as indicated within the present work.

## 5. Conclusions

In this work, we have investigated the effect of local phase morphology on deformation process of a dual phase Ti-6242 using combined *in-situ* micropillar compression and HR-EBSD, together with post-mortem TEM. Micropillars containing two similar α-orientations but different β laths (i.e. inclined and vertical) were compressed. The mechanical performance and local variation in deformation within the pillars indicate that the presence, structure and properties of the β-phase can significantly alter mechanical properties within a dual-phase titanium microstructure. This was seen in the load-displacement response, as a change in apparent yield point, local hardening rate and strain rate sensitivity. It is likely that these changes are due to the role of both the α/β interface and the β phase, which changes the activity of slip systems within these micropillars. This is evidences through a change in the slip steps seen on the side of the pillars, as well as the homogeneity of the GND density and stress state within the pillars. The pillar containing vertical β showed significant hardening associated with the pair of slip bands that operated, whereas the β-inclined pillar showed a significantly more homogeneous deformation (consistent with less hardening). These results indicate that the β-phase has a significant opportunity to modulate the stress state, strain rate sensitivity and load-shedding of dual phase titanium microstructure which is likely to play an appreciable role in stress accumulation and the associated load-shedding which is thought to be important for dwell fatigue of Ti alloys.

**Acknowledgement**

We are grateful to the Engineering and Physical Science Research Council for funding through EP/K028707/1 and the HexMat programme grant (EP/K034332/1). Further details of the HexMat grant can be found at http://www.imperial.ac.uk/hexmat. TBB would like to acknowledge funding from the Royal Academy of Engineering Research Fellowship scheme.

We also would like to thank Professor Fionn Dunne (Imperial College) for helpful discussions on cold dwell fatigue and local deformation mechanisms of Ti alloys, and Mr. Damian Frey (EMPA) for his help with the Alemnis indenter.

**Supplementary online information**

The individual compression data associated with this article can be found in the online version at (Zenodo DOI will be added at article acceptance).